# Quantitative error analysis of back-stripping based models: case study from Po Delta (northern Italy)

Vitagliano E.[1,3*], D'Ambrogi C.[2], Spassiani I.[1] and Di Maio R.[3]

[1]*Istituto Nazionale di Geofisica e Vulcanologia (Rome, Italy)*
[2]*Istituto Superiore per la Protezione e la Ricerca Ambientale (Rome, Italy)*
[3]*Dipartimento di Scienze della Terra, dell'Ambiente e delle Risorse, Università degli Studi di Napoli Federico II (Naples, Italy)*
[*]*eleonora.vitagliano@ingv.it*

**Abstract**
Numerical back-stripping procedure is crucial for understanding the geological mechanisms of basin formation or for reconstructing the palaeo-bathymetry in the oceanic regions. However, the importance of errors associated with data acquisition, processing and interpretation is often underestimated. These errors, which can impact the final results, are not part of the computational workflow, although they often affect the model parameters with large uncertainties. In this study, we have qualitatively classified and quantified all the main errors affecting the workflow of the back-stripping technique using linear interpolation and combinatorics. We found that the errors influence different model parameters, some of which have an equiprobability of occurrence, while others are characterized by an intrinsic probability. We applied the proposed method to the Po Delta in northern Italy, historically influenced by anthropogenic and natural subsidence. By studying a 2D geological section characterized by thin Holocene sedimentary successions, we identified 12 sources of error that fall into three basic categories: geometry of the model layers; distribution of lithologies and petrophysical properties; past depositional environments. We then assessed the error ranges and their probability of occurrence. The study shows that the errors can vary significantly from metre- to millimetre-scale, defining the magnitude and distribution of each error source, which is essential for interpreting model results and assessing related uncertainties. It establishes a workflow for future uncertainty management and aims to enhance open-source tools based on the back-stripping procedure.

## 1. Introduction

In the field of quantitative geology, subsurface models play a crucial role in describing the complexity of geological structures and in understanding the associated processes and dynamics. These models integrate various data sources, such as geophysical surveys, borehole data, and surface maps, to provide representations of subsurface geological units, mineral distributions, tectonic structures, etc. However, the accuracy of these models is inherently limited due to uncertainties in data acquisition, processing, and interpretation. Consequently, error analysis in geological modelling is essential to quantify uncertainties and assess model reliability (Caers, 2011). In fields such as resource exploration, environmental management, and geotechnical engineering, where decision-making depends on the reliability of subsurface predictions (e.g., Wellmann & Caumon, 2018), it is essential to identify sources of error and understand how they propagate. The complex nature of geological systems poses unique challenges for error analysis, requiring advanced computational and statistical techniques to manage uncertainty. Variations in data quality, such as limitations in seismic resolution or incomplete datasets, can significantly impact the model's predictive capabilities. In addition, simplifications inherent in modelling algorithms, such as assumptions about linearity in geophysical inversion or stratigraphic continuity, can introduce further bias. Modelling errors can also arise from the precision of the instruments used for data acquisition, or from the sampling, transformation or normalization of the input data (e.g., Davis, 2002; Press et al., 2007; Aydin, 2013). They may even depend on the natural variability of the analyzed features or simplifications due to the model's computational assumptions (e.g., Journel & Huijbregts, 1978; Beven, 2009; Chiles & Delfiner, 2012). Temporal uncertainty can affect the age assigned to geological strata or, more broadly, the environmental factors that influence the analyzed parameters (e.g., Price, 1997; Steiger & Jäger, 1977). Furthermore, interpretative biases and subjective errors can affect several variables in the modelling calculations, largely depending on the knowledge of the modeler (e.g., Krumbein & Graybill, 1965; Bond et al., 2007; Madsen et al., 2022). While the calibration process can help reduce errors, it often focuses on a limited number of model parameters (e.g., Müller & Butler, 2014). Therefore, error analysis in geological models addresses a wide range of issues, requiring proper identification and quantification of errors before uncertainties can be reduced.

Among the numerical methods applied in the geological models, the back-stripping technique has been widely used to quantify basin subsidence history, sedimentation rates and tectonic subsidence (Watts & Ryan, 1976; Steckler & Watts, 1978; van Hinte, 1978; Falvey & Deighton, 1982). In particular, in the last decades, numerous studies have applied this technique, focusing on the understanding of basin formation mechanisms (e.g., Royden & Keen, 1980; Hegarty et al., 1988; Berra & Carminati, 2010; Khan & Abdelmaksoud, 2020) or the assessment of petroleum systems at regional and local scales (e.g., Welte et al., 1997; Kauerauf & Hantschel, 2009; Schenk et al., 2012; Tawfik et al., 2023), as well as for the reconstruction of paleo-water depths, especially in environmental studies related to oceanic or Arctic regions (e.g., Kjennerud & Sylta, 2001; Roberts et al., 2003; Lasabuda et al., 2023). The recent availability of Matlab-compatible open-source codes to achieve these goals indicates a significant interest in accessible procedures to obtain the aforementioned results (e.g., Lee et al., 2020; Colleoni et al., 2021). However, a systematic investigation of the main errors associated with data acquisition, processing and input or model parameters has not yet been carried out. In fact, the standard approach to dealing with errors in input data is to test different scenarios and to calibrate calculated values against observed data. Some researchers have approached error analysis by focusing on selected parameters, such as those associated with porosity reduction mechanisms or those related to the variations in water depth and sea level (e.g., Watts & Ryan, 1976; Gallagher, 1989). Other authors have quantified errors affecting the Earth's surface due to isostatic loading effects (e.g., Zhou et al., 1993; Roberts et al., 1998) or introduced erosion thickness into their models to assess the timing of maximum burial (e.g., Baig et al., 2019). However, most of the cited studies have focused on basins characterized by sediment thicknesses and ages of several thousand meters or hundreds of millions of years, respectively, where errors of a few meters in thickness or a few million years may be negligible. Conversely, only a limited number of studies have applied the back-stripping procedure to areas characterized by thin sedimentary successions, where a few tens or hundreds of meters of sediments have accumulated over relatively short time intervals (e.g., Steckler et al., 1999; Maselli et al., 2010). In these geological contexts, a thorough error analysis could be crucial, as even small discrepancies in sediment thickness or age can have a significant impact on the final results, thus requiring a careful treatment of these errors to better understand their distribution and significance.

This paper presents a comprehensive error analysis in the context of geological modelling, specifically aimed at evaluating natural subsidence rates using a back-stripping procedure. By studying a 2D geological section traced in the Po Delta area (northern Italy) and characterized by thin Holocene sedimentary successions, we identified 12 sources of error related to data acquisition, processing, and interpretation. These sources were then integrated into the modelling workflow, concerning both input and model parameters. Our qualitative and quantitative analysis is particularly applicable to sites with high data uncertainty, such as the case study described here, and is particularly valuable in defining the magnitude and distribution of specific errors. Understanding these factors is crucial for interpreting the model's results and for calculating the uncertainties associated with each component of the model. Thus, the proposed study establishes a solid workflow for future efforts in uncertainty management and aims to improve the open-source tools based on the back-stripping procedure.

## 2. Method

### 2.1 Back-stripping technique

The back-stripping technique is widely used in sedimentary basin analysis, petroleum system modelling and tectonic studies (e.g., Van Hinte, 1978; Berra & Carminati, 2010; Lee et al., 2020; Aladwani, 2021). More recently, it has been used to reconstruct paleo-bathymetry in continental margins and the Arctic Ocean (e.g., Kjennerud & Sylta, 2001; Lasabuda et al., 2023). The back-stripping procedure starts from the current sediment thickness and works backwards in time, systematically removing layers while allowing for sediment compaction and isostatic adjustments due to sediment and water loads. In modelling studies focused on the evaluation of subsidence rates, the total subsidence (BD) at each time step-representing the depth reached due to sedimentary, loading and tectonic processes (e.g., Allen & Allen, 2005)-is calculated using the following formula:

$$BD = DT + PSL + PWD, \quad (1)$$

where DT is the decompacted thickness, PSL is the paleo-sea level and PWD is the paleo-water depth at each time step. Additionally, it is possible to determine the model base depth due to tectonic processes alone by modifying Equation (1) as follows:

$$BD_t = BD + LC, \quad (2)$$

where $BD_t$ is the model base depth resulting from tectonics and LC is the vertical shift associated with the loading component. It is important to note that in the proposed study, depths measured below the mean sea level (msl) are considered positive. Consequently, an increase in the model base depth over time, such as that caused by subsidence, is treated as a positive variation, whereas a decrease in the model base depth over time, such as that caused by uplift, is treated as a negative variation.

To perform back-stripping analysis and calculate subsidence rates, the following input data are required: horizon depth, horizon age, petrophysical properties of the lithotype, elastic properties of the lithosphere, paleo-water depth, and paleo-sea level. Based on these data and additional calculations, the following model parameters can be obtained: layer thickness, time interval of the layers, decompacted thickness, isostatic correction, paleo-water depth, and paleo-sea level. Some input data, such as paleo-water depth and paleo-sea level, are incorporated directly into Equation (1). In contrast, other data undergo additional calculations before being used as model parameters, such as thickness measurements and time intervals derived from horizon depth and age differences, respectively. As the focus of this research is on error analysis, these data and parameters will be examined in detail here, while the final subsidence rates will be considered and discussed in future research.

### 2.2 Qualitative aspects of proposed error analysis

According to the input data presented in Section 2.1, we categorize the error-generating features into four main groups or sources (Fig. 1).

The first group includes errors related to the horizon depth, which directly affect the model layer thickness or geometry. These errors can arise from both interpretation issues, for example where multiple depth interpretations exist for the same stratigraphic horizon (Fig. 1a), and from the procedures used to acquire the data, such as data resampling (Fig. 1b). Systematic errors can also arise from the size of the cursor used to pick horizons, particularly if its size is significant relative to the vertical and horizontal scales of the depth profile (Fig. 1c). Modeler-related errors can occur when the interpretation of a depth horizon is dependent on, or independent of, a dataset. For example, if a modeler selects a horizon on a seismic line, the point selected may not accurately reflect the true stratigraphy of the horizon, even if the interpretation is based on seismic data. In addition, a new dataset may lead to revised depth horizon interpretations.

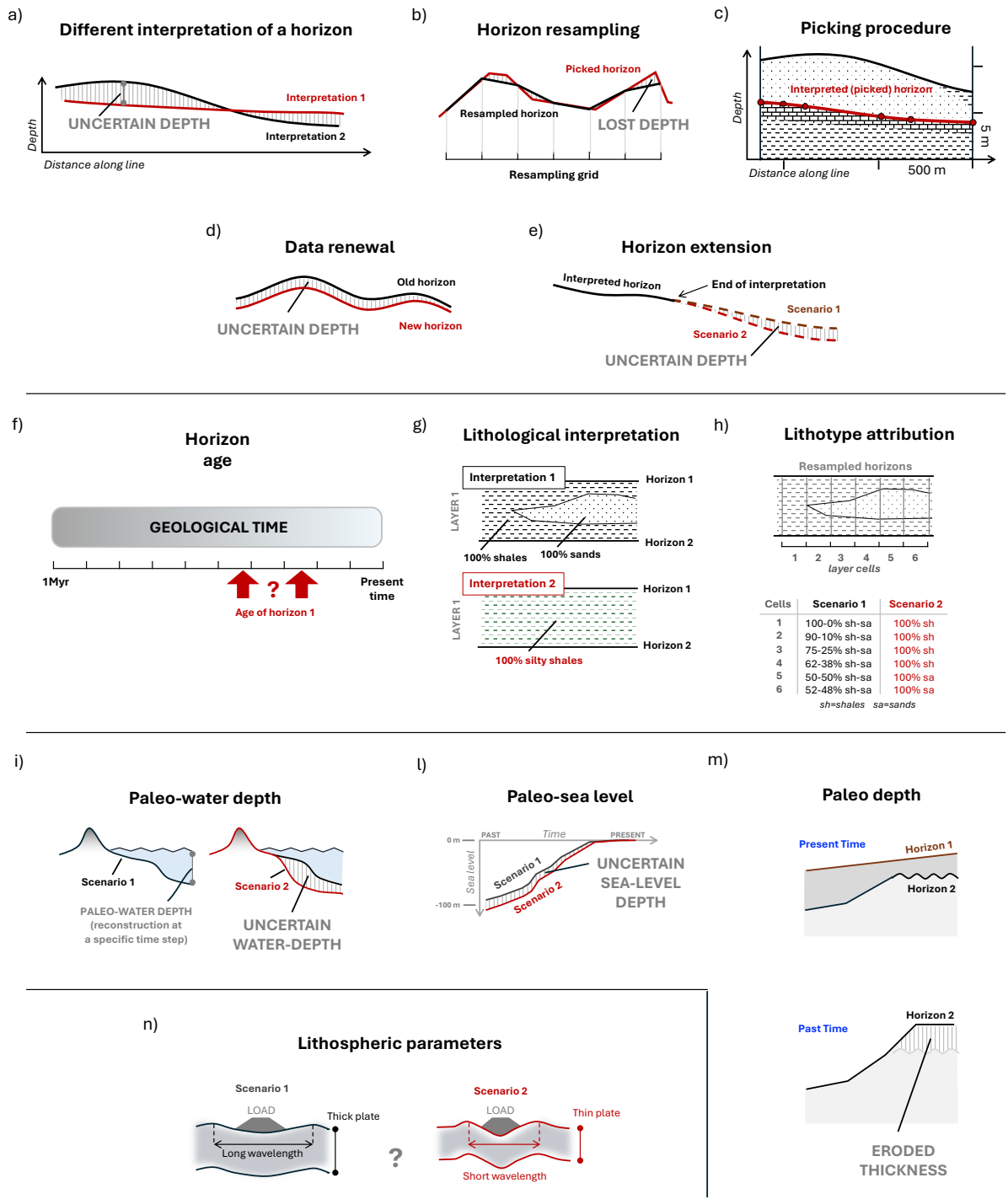

*Figure 1. Main errors associated with input data and model parameters of the back-stripping procedure. Note that the error sources refer to: a)-e) layer geometry, f)-h) stratigraphic-sedimentological aspects, i)-m) depositional-environmental context, and n) geodynamic features.*

If a critical review of old data with respect to new data is not performed properly, the data update may introduce new errors in depth values, thereby increasing uncertainty (Fig. 1d). Finally, in some cases where seismic horizons are not visible on the seismic profile, horizons can be drawn based on geological assumptions. This is often the case in the depocenter of the basins, where the stratigraphic horizons are geologically expected to be plane-parallel to the basal surface bounding the basin (Fig. 1e).

The second source of error relates to the stratigraphic-sedimentological aspects that affect the age of the model layer and the decompacted thickness in relation to the depth-porosity trend. The first aspect, i.e. the horizon age, is often subject to uncertainty due to the methodologies and datasets used for its identification (see Fig. 1f). In addition, these methods and datasets can vary depending on the geo-disciplinary perspective of the researcher studying the issue (e.g., geochronology versus stratigraphic correlation). This type of uncertainty also leads to discrepancies in other aspects, such as the determination of paleo-sea levels and paleo-water depths. Another source of error is the sediment and rock distribution within the model layers. This can vary according to the modeler's knowledge, experience, objectives, and available datasets. Indeed, multiple sedimentological interpretations can exist for the same geological profile (Fig. 1g), and a single layer can be associated with different lithotype assignments (Fig. 1h). Within a layer cell, a lithotype defines the proportion of one type of lithology relative to another (e.g., sandstones, shales, limestones, etc.). This attribution is crucial in geological modelling as it determines the ranges of petrophysical properties required for further calculations, such as thermal conductivity, hydraulic conductivity, and permeability. Although this process is essential, lithotype attribution is rarely based on quantitative studies. Typically, detailed information is gathered at a local scale and then extrapolated based on additional rules or assumptions influenced by the modeler's expertise and understanding. Finally, commercial geological modelling tools often impose fixed percentages of lithotypes that can be attributed to model layers (e.g., 0-25-50-75%) and limit the number of mixed lithologies (e.g., a maximum of two). If the chosen attribution lacks a calibration phase that aligns some petrophysical parameters with accurate lithology data, it can introduce further errors.

The third source of error focuses on depositional-environmental aspects, specifically paleo-bathymetry and paleo-sea level (Figs. 1i-l). This category relies on the modeler's knowledge and experience, data availability and advanced research. It draws on geological information at local and regional scales, including seismic profiles, micro-paleontological studies, literature on relative and eustatic sea-level changes, and paleo-environmental studies. In addition, this category includes uncertainties related to eroded thickness (Fig. 1m). Although this type of error relates to layer geometry, it is included here because it concerns the paleo-geometric reconstruction that often employs paleo-environmental data or proxies that are sensitive to environmental conditions, such as vitrinite reflectance, which is a terrestrial maceral and indicates maximum burial depth combined with thermal conditions.

The final category relates to the properties of the lithosphere (Fig. 1n), such as mantle density, elastic thickness, and flexural wavelength. These properties are essential for computing isostatic corrections, for example according to the flexural isostatic deflection model proposed by Turcotte and Schubert (2002). The values assigned to these variables are mainly influenced by the modeler's geodynamics expertise and understanding of the study area, and by the availability of regional studies in the literature.

In summary, Figure 1 illustrates the main errors associated with the proposed modelling. However, it does not provide a comprehensive list of all potential errors in geological modelling, although some errors not considered here may be related to those described in the figure. For example, errors resulting from seismic data processing or time-depth conversion may resemble those related to the layer geometry shown in Figs. 1a and 1c, respectively, although the uncertainties linked to these error sources are not addressed in this study.

## 2.3 *Quantitative aspects of proposed error analysis*

The quantification of errors is achieved step by step following the qualitative categories described in Section 2.2. Specifically, each type of error is quantified in terms of a range within which data can be found, from the minimum to the maximum data values. These lower and upper bounds of the error range are determined by using standard procedures implemented in Matlab, mainly based on resampling techniques, combinatorial methods and automatic pixel-counting tools (e.g., Kenneth, 1999; Phan, 2010; Siogkas, 2013).

The resampling techniques, including both regular (constant) and irregular resampling rates with respect to the original sampling, are based on linear interpolation and herein applied to identify the first source of errors (those relative to the horizon depths). This process returns the interpolated depth values of a one-dimensional function at specific query points. The latter correspond to either regularly spaced or irregularly spaced sampling points/model nodes along the line distance.

Regarding the combinatory methods, the techniques of arrangement with repetition and simple arrangement are used here to identify the possible errors, which are treated as resultant combinations. In particular, the first operation involves calculating the number of ways to arrange $n$ elements (or values) in $m$ positions. The elements can be repeated, and the operation returns the $m^n$ combinations. In our case study, it is applied to evaluate the vertical picking error, where $n$ elements are given by the maximum and minimum vertical error amounts, and $m$ positions are given by the involved horizons. The second operation combines the number $n_1$ of elements in a first group with the number $n_2$ of elements in a second group, without repetition. It returns $n_1 * n_2$ combinations and is used here to evaluate the horizontal picking error. In this case, $n_1$ and $n_2$ correspond to the maximum and minimum horizontal errors with respect to a horizon point (first group) and to the consecutive one (second group).

Finally, a pixel-counting procedure that automatically recognizes the colours associated with each lithology from geological section images is used here to manage the error associated with the lithological interpretation and lithotype attribution. This operation improves the data-driven selection and avoids other procedures that select the lithotype contents based on fixed percentages (e.g., sands 25% - shales 75%). It is obtained as a combination of a step-by-step procedure, starting from the conversion of the colour values into a matrix containing the number of pixels related to Red, Green and Blue (RGB) colours, then selecting the frequency of specific colours and finally defining the pixel counting. The procedure has been implemented in the MATLAB language (Danz, 2020).

## 3. Application of error analysis to the Po Delta

The study area is located in the Po Delta (northern Italy), a coastal region involved in complex interactions between natural processes and human activities (Fig. 2a). This area is characterized by a highly dynamic environment, where the balance between sediment deposition and land subsidence, also influenced by climate effects, affects the delta's morphology (e.g., Simeoni & Corbau, 2009; Antonioli et al., 2017). Historically, the area has faced significant land subsidence due to the extraction of underground fluids, such as water and natural gas, and the consequent compaction of shallow sedimentary layers (e.g., Bondesan & Simeoni, 1983; Carminati & Martinelli, 2002). Currently, the delta experiences natural subsidence rates of a few millimetres per year, which increase to the east, where fine-grained sediments, such as silts and clays belonging to the inter-distributary channels and the lower delta plain, are mainly involved in consolidation processes (e.g., Teatini et al., 2011; Gambolati & Teatini, 2015).

For the purposes of this study, a comprehensive dataset of geological information on the Po Delta area has been collected and is presented in Table 1. The dataset is mainly related to the 1:25.000 scale geological survey carried out for the production of the geological Sheet 187 Codigoro and the associated subsurface geological map (Servizio Geologico d'Italia, 2009a and 2009b), along with the accompanying notes (Cibin & Stefani, 2009). As detailed in Supplemental Information S1, Sheet 187 provides both shallow and deep geological sections used for the error analysis. Both sections refer to the northern part of the geological map, run in the same direction (WWN-EES) and have a total length of 26 km (see Fig. 2b). They cut 20 m of the latest Pleistocene-Holocene sediments deposited in marine and deltaic environments.

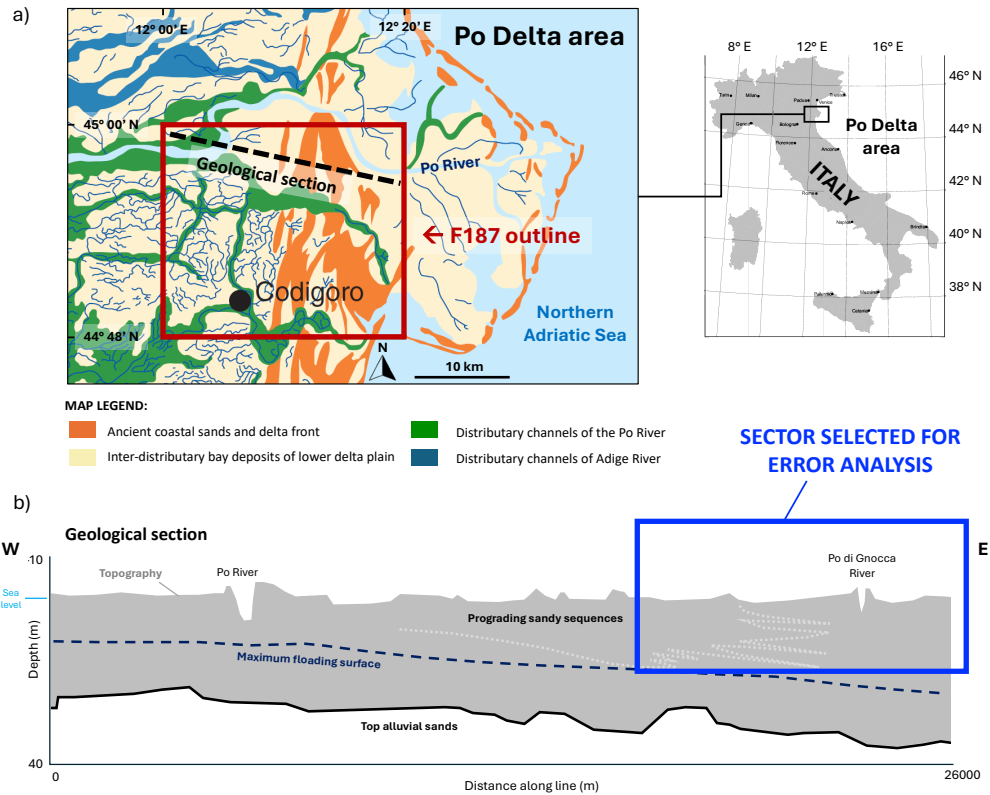

*Figure 2. Po Delta area: a) geological map and b) 2D section used in the back-stripping analysis. Note that the map is modified from the Sheet 187 Codigoro, Geological Map of Italy at the scale 1:50.000 (Servizio Geologico d'Italia, 2009a).*

Within these sections, the eastern part of the Holocene succession contains sandy sediments from prograding bodies interbedded with pro-delta muddy sequences. This complexity results in considerable variability in paleo-environmental reconstruction and lithological distribution, making it a suitable subject for error analysis. In addition, there is a notable discrepancy between the input data due to the different disciplinary backgrounds of the data provenance and the different methodologies employed for data acquisition, processing, and interpretation. This discrepancy contributes to the analysis of error sources.

| MODEL PARAMETER | DATABASE | DATA SOURCE |
|---|---|---|
| Horizon depth | Line AA' of Geological Map 187 (surface sheet) Line CC' of Geological Map 187 (subsurface geology sheet) | Servizio Geologico d'Italia, 2009a and 2009b |
| Topography | Cumulative subsidence map (period: 1900-2015) | Corbau et al., 2019 |
| Horizon age | Chronostratigraphic scheme Chronological evolution of Po Basin and Adriatic Sea Measures of radiocarbon age | Cibin & Stefani, 2009 Amorosi et al., 2016; 2017 Cibin & Stefani, 2009 |
| Lithotypes | Lithological boreholes (187-S4 and 187-S3)<br>Penetrometer tests: 187080U50 (4 sites),<br>187020U50 (4 sites), 187NEU5 (7 sites)<br>Environmental and sedimentological data<br>Sedimentological and lithological study<br>Geological Maps: 1:50,000 F. 187; 1:100,000 F. 77 and F. 65 | Cibin & Stefani, 2009<br>Geoportal of Emilia-Romagna region<br>Cibin & Stefani, 2009<br>Cibin & Stefani, 2009<br>Servizio Geologico d'Italia, 1954; 1967; 2009a and 2009b |
| Petrophysical properties | Porosity<br>Bulk density shallow sequences | Sclater & Christie, 1980<br>Baldwin & Butler, 1985<br>Tornaghi & Perelli Cippo, 1985<br>Maselli et al., 2010<br>Cortellazzo & Simonini, 2001 |
| Lithosphere elastic properties | Elastic thickness, flexural wavelength<br>Mantle density | Carminati & Di Donato, 1999<br>D'Agostino et al., 2001<br>Carminati et al., 2010<br>Maselli et al., 2010<br>Gvirtzman et al., 2016 |
| Glacial rebound | Isostatic correction due to ice melt | Carminati & Di Donato, 1999<br>Carminati et al., 2003 |
| Paleo-sea level | Paleo-sea level trend | Lambeck et al., 2011; 2014<br>Geological Map of the Italian Sea 1:250.000 NL-33-7-Venice (Servizio Geologico d'Italia, 2010) |
| Paleo-bathymetry | Core1 and CoreS1 micro-paleontological study<br>Micro-paleontological study on 187-S1<br>Paleo-coast lines<br>Environmental and sedimentological data<br>Pro-delta and shallow marine bathymetry<br>Dune system profile<br>Foraminifer depth distribution | Rossi & Vaiani, 2008<br>Cibin & Stefani, 2009<br>Antonioli et al., 2009<br>Cibin & Stefani, 2009<br>Correggiari et al., 2005<br>Donadio et al., 2017<br>Aiello et al., 2012 |

*Table 1. Dataset used to perform the error analysis in the Po Delta area. Data are grouped according to the model parameters and data providers.*

## 4. Results

### *4.1 Errors related to horizon geometry*

The available datasets provide two distinct geological sections of the same profile (Fig. 3), representing the shallow and deep interpretations of Sheet 187 (hereafter referred to as the shallow and deep sections, respectively). These lines overlap in the upper strata consisting of Holocene sandy and clayey deposits from the deltaic plain, delta front, and coastal plain. Two sets of horizon depths and relative line distances are obtained by digitizing the section images in the selected sector of the study site (see also Fig. 2b) over a length of 8,570 m. The digitized horizons share the same geological significance in both sections, preserve the geometry of the original horizons and remain synchronous throughout the entire sector length. Specifically, the horizons correspond to the topography (Horizon 1) and the top of the prograding stack (Horizon 2).

Moreover, to quantify the error due to the two interpretations, all the digitized horizons (original horizons) were resampled at a rate of 1 mm. This rate was chosen to match the precision of the horizon depths, which are given in meters with three decimal places. For each interpretation, the thickness of the upper layer (Layer 1) is calculated as the difference between the top of the prograding stack and the topography (Figs. 3a and 3b).

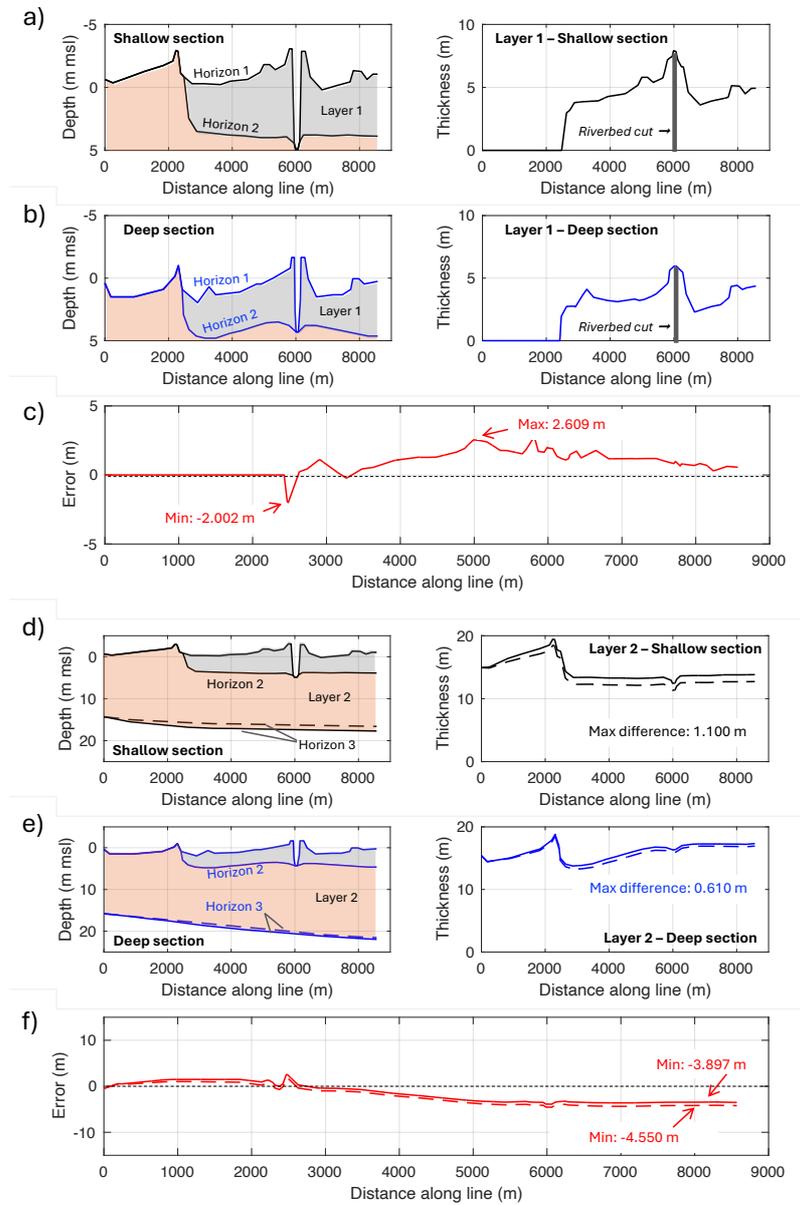

*Figure 3. Errors due to different interpretations of section horizons (a-c) and due to arbitrary extension of undefined horizons (d-f). Horizons 1, 2 and 3 correspond to topography, top and bottom of the prograding stack, respectively. It is important to note that in the a)-c) plots, the riverbed intersecting the lines at about 6000 m along the line distance has been removed in the layer thickness plots on the right to better visualize the error trend. In d)-f) plots the horizon extension is depicted according to two scenarios: scenario 1 is shown with a continuous line, while scenario 2 is represented with a dashed line. See the text for a detailed explanation.*

In particular, the error associated with the layer thickness of the two sections is determined for difference of layer thickness and is displayed in Figure 3c. The plots show that the two thicknesses follow qualitatively the same trend after the riverbed cut, whereas before this point, the shallow interpretation marks a slightly higher thickness than the deep one. The error is quantified within a range of -2.002 and 2.609 m, with a mean of 0.845 m, which is not entirely negligible considering the relatively thin layers accounted in this analysis (less than 10 m). Therefore, in the case of different interpretations of the horizons, it would be a good practice to quantify the associated uncertainty that may affect the results. A method to pursue this issue is the Min-Max approach, where the absolute uncertainty is estimated as half the difference between the minimum and maximum in the dataset. Another source of error related to the horizon depth is the arbitrary extension of key horizons across the entire length of the sector. Specifically, the base of the prograding stack (Horizon 3) is not traced in the original section images. Thus, the geometry of this horizon is not fixed and may vary if reliable extension criteria are applied. In this case study, two criteria were used to extrapolate the horizon: maintaining the geometry of the overlying layer (scenario 1) or the underlying layer (scenario 2). This assumes that sedimentary bodies deposited under similar energy conditions and environments will also have similar shapes. As mentioned above, the arbitrarily extended horizons are first resampled, and then used to calculate the thickness of Layer 2 as the difference between the base and the top of the prograding stack horizons (Figs. 3d and 3e). Finally, the error is evaluated as the difference between the two calculated layer thicknesses (Fig. 3f).

It is worth noting that two sources of uncertainty are embedded in this case: one again due to the different interpretation of the horizon (shallow and deep sections), and the other related to the horizon extension (scenario 1 and 2). Figures 3d, 3e, and 3f clearly show that the final error is mainly influenced by the first source of uncertainty (different horizon interpretations). The discrepancy increases in absolute value as the distance along the line rises, reaching 3.897 m and 4.550 m for scenarios 1 and 2, respectively. Instead, the proposed scenarios for extending Horizon 3 appear to be less relevant, as the maximum differences observed in the shallow and deep interpretations are 1.100 m and 0.610 m, respectively.

The error related to the picking procedure is assessed by comparing the cursor size of the digitizer tool used (GetData Graph Digitizer, 2024) with the vertical and horizontal scales of the original geological sections. Specifically, for the vertical scale, the cursor dimensions correspond to 0.55 m in the shallow section and 1 m in the deep section. These values indicate the range of uncertainty in the vertical positioning of the represented depths at the original sampling points. It can therefore be assumed that each depth value corresponds to the original picking depth ± 0.28 m and ± 0.50 m in the shallow and deep sections, respectively. Similarly, for the horizontal scale, the cursor size corresponds to 70 m in the shallow section and 56 m in the deep section. These values represent the range of uncertainty in the horizontal positioning of the depths shown. It can therefore be assumed that the distance value along the line corresponds to the original picking distance of ± 35 m and ± 28 m in the shallow and deep sections, respectively.

Moreover, additional operations are required to transfer the error from the horizon depth to layer thickness and to calculate the maximum possible changes in the layer thicknesses. Specifically, in the case of a vertical picking error, the maximum increase or decrease in the relative thicknesses can be determined by simultaneously adjusting the depth values of all the horizons at their original sampling points by the same error amounts (±0.28 m for shallow sections and ±0.50 m for deep sections). The excess or deficit in the layer thicknesses can therefore be derived by combining the two error values over three horizon depths. The results consist of eight possible combinations (Fig. 4a). From a purely mathematical point of view, if dz is half the dimension of the cursor, we actually have 3 horizons and, for each of them, 2 possible picking depths (i.e., ±dz), giving $2^3 = 8$ permutations by repeating the combinatory method introduced in Section 2.3 (with m = 3 and n = 2). Considering that all three horizons can have the same direction of displacement (upward or downward), in terms of layers we have that in two out of the eight cases, there is a thickness variation of ±0.55 m in both layers. In four of the eight cases, the thickness variation affects only one layer (either Layer 1 or 2). In the remaining two cases, there is no thickness variation, although the horizon depths are affected by the error.

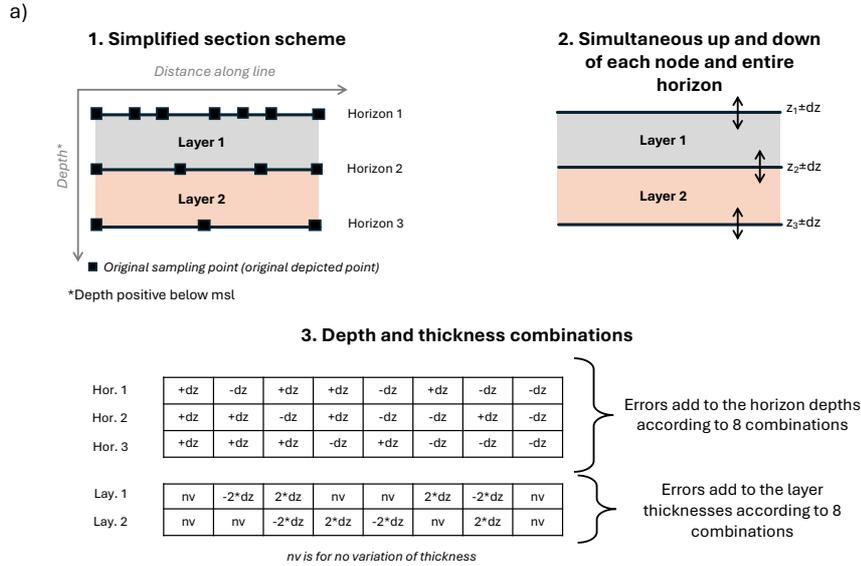

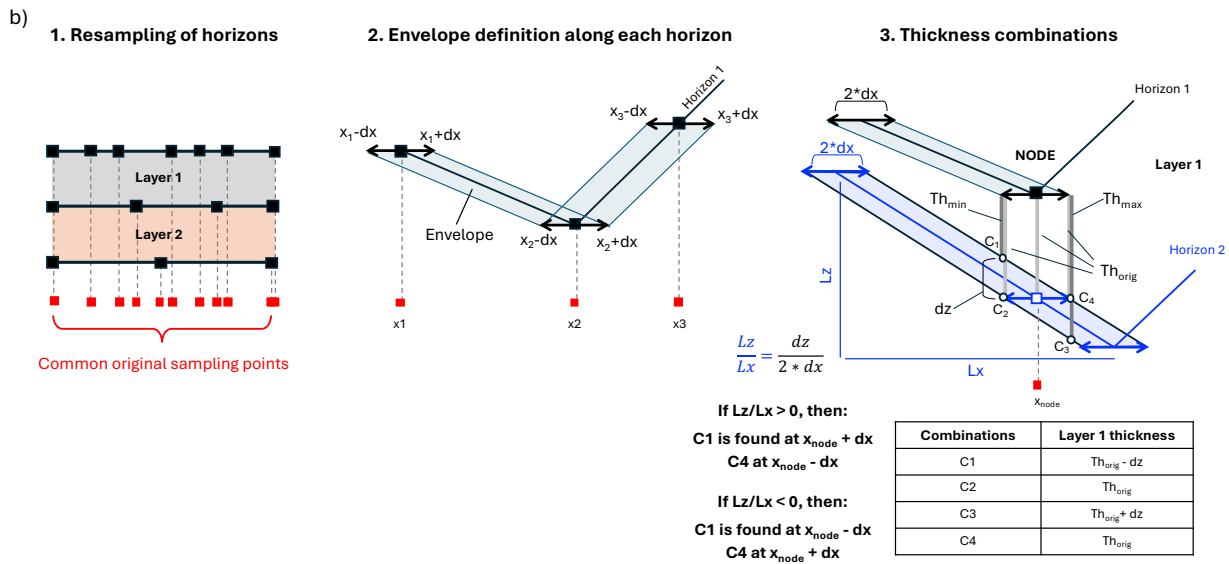

Figure 4. Procedure for calculating the errors due to a) vertical and b) horizontal picking.

For the horizontal picking error, the procedure for calculating the thresholds differs from the previous case. First, the sampling points along the line distance of the three horizons were combined into common sampling points (Fig. 4b). This was done by concatenating and sorting the line distance values of each horizon, removing duplicates. A one-dimensional linear interpolation function was then applied to obtain interpolated depth values at specific query points of the common sampling points for each horizon. This resulted in a total of 50 original sampling points for the shallow section and 47 for the deep section. After establishing the common sampling points, we introduced a combination technique to find the envelope that accounts for the horizontal error in each node ($\pm$ dx) and connects two consecutive nodes within the same horizon. This envelope is defined by four combinations, resulting from two possible horizontal shifts between two consecutive nodes, as illustrated in Figure 4b. Within this envelope, there are four possible thickness combinations: two correspond to the original thickness, one represents the maximum thickness, and one indicates the minimum thickness. The geometry of the horizons determines the maximum and minimum values and, in particular, the variation of the original thickness depends on the dip of the horizon where the examined node is projected, as well as the original thickness at that node and the horizontal error. To generalize, the error region is framed within the intersection between any vertical line, and the parallelogram, where the two sides are respectively equal to the cursor dimension and the nodes' distance along the line, while the angle is the horizon slope (dz in the third panel of Fig. 4b). In the specific case of the Po Delta area that we are focusing on, the resulting errors vary along the line, ranging from -0.61 to 1.53 m in the shallow section and from -0.2 to 0.31 m in the deep section, the latter appearing to be much less affected by this source of uncertainty.

When comparing this error category with the others mentioned above and below, it is clear that the occurrence of a thickness variation is associated with a specific probability. In contrast, for the other types of error, such as differing geological interpretations and arbitrary extensions, all error values are treated as equally probable (each assigned a probability of 1) within the maximum and minimum thresholds.

The error associated with resampling, which is a necessary procedure during model development to optimize computation, arises from the use of a constant resampling rate across all horizons. In particular, the original sampling points, obtained from the picking of the section images, vary for each horizon and depend on the geometry, since these points are selected at the inflection points and edges of the horizons. When a resampling rate is introduced, the original and resampled points may not be perfectly aligned; this misalignment occurs either because the new sampling step differs in size from the original or because it is not a multiple of the original sampling interval. To assess the errors, we evaluated four constant resampling rates: 1 mm (rate-1), 1 m (rate-2), 10 m (rate-3) and 100 m (rate-4). As rate-1 replicates the original digitized horizon, it serves as a reference for the other rates, allowing us to calculate the depth losses incurred due to the resampling procedures. This calculation involves the application of linear interpolation to the query points that are aligned with the various rates. The depth losses were then analyzed to determine the differences between the resampled horizons and the reference horizon (at the same nodes).

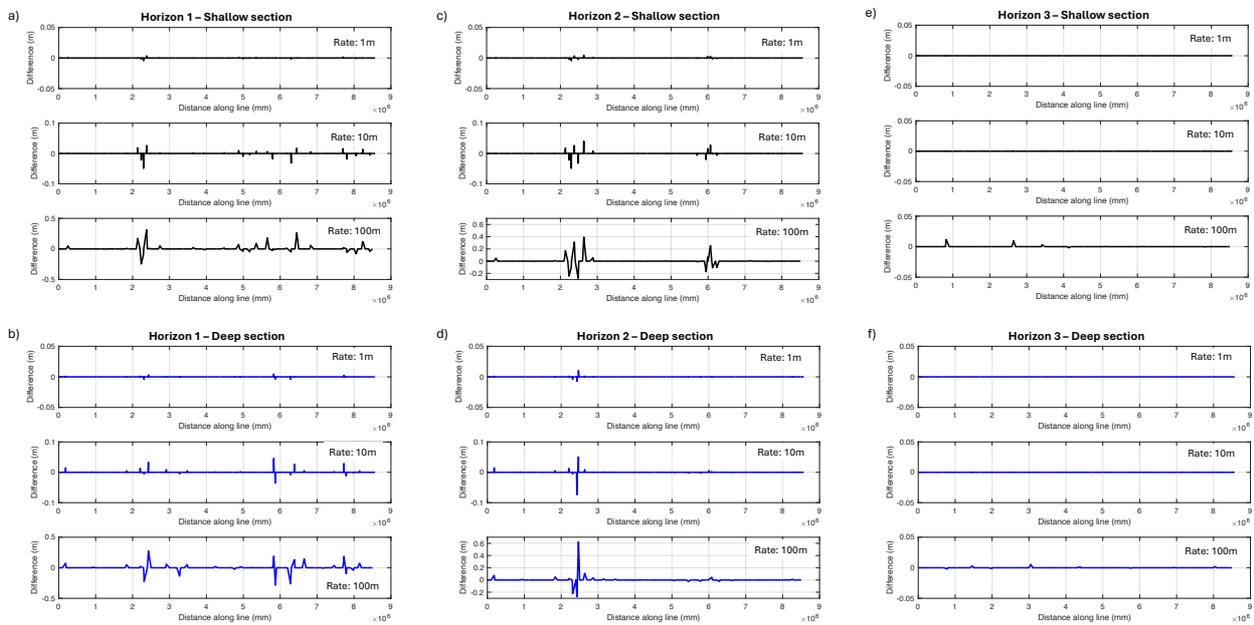

*Figure 5. Errors due to differences in resampling rates across all horizons. Specifically, all sketches display the differences (in meters) between the horizons sampled at rates of 1 m, 10 m and 100 m compared to the reference rate of 1 mm (see text for further details). Note that the vertical scales in the a)-d) plots vary between the upper, middle and lower sketches to highlight the differences.*

Figure 5 shows the errors produced by the different resampling rates, highlighting that as the resampling rate increases, so does the depth difference (i.e. error or loss of depth relative to the original depth). For example, between 2,000 and 3,000 m of line distance, Horizon 1 experiences depth losses of up to ±0.01 m at a resampling rate of 1 m in both shallow and deep sections (Fig. 5a and 5b). This error increases to ±0.05 m at a resampling rate of 10 m and reaches ±0.25 m at a rate of 100 m. In the case of Po Delta area, Horizon 3 also seems to be the least affected by this source of uncertainty, but no general trend can be deduced with respect to the depths of the horizons: the picking number in the resampling procedure is generally related to the complexity in the geometry of the horizons considered. For this reason, this analysis is not definitive in defining a resampling method for modelling natural subsidence. However, in our case study, it implies the selection of a 10 m rate to optimize the computational processes. This results in 858 nodes along the sections, corresponding to 857 computational cells within the 2D models. Finally, it is also worth highlighting that there are no significant differences when comparing the results obtained for the two interpretations. In fact, the shallow case is slightly more affected by this source of error than the deep case, but in general, the discrepancy is negligible.

### *4.2 Errors due to uncertain horizon ages, lithotypes and paleo-environments*

The information used to characterize the geological sections reveals discrepancies in the ages of the horizons (see also Tab. 1). For instance, the notes accompanying Sheet 187, which describe the datasets collected for the local geological map (Cibin & Stefani, 2009), date Horizon 2 at 0.4 thousand years (kyr) and Horizon 3 at 3.5-4.5 kyr. Conversely, the studies by Amorosi et al. (2016; 2017), which focus on sequence stratigraphy in the Po Delta area and its surroundings, provide ages of 0.8 kyr for Horizon 2 and 5 kyr for Horizon 3. Consequently, the age ranges for horizons 2 and 3 are from 0.4 to 0.8 kyr and from 3.5 to 5 kyr, respectively. Finally, the relative error ranges from 0 to 0.4 kyr for Horizon 2 and from 0 to 1.5 kyr for Horizon 3, indicating that it increases with the age of the horizons. This result is to be expected given the different techniques employed to determine chronology or age and the inherent uncertainties in their results.

Regarding the lithological distribution presented in the shallow and deep section images (Fig. 6a), it shows minor differences in the contents of sands, sand-silts, and clay-silts within the section layers. A notable difference is the introduction of a third lithotype category, i.e. sand-silts, found in the deep section. In addition, the sandy prograding stack extends more westward in the shallow section than in the deep section.

To quantify the lithotypes and define the relative porosity-depth curve for each layer, we first grouped the nodes using a regular grid with a mesh size of 1,000 m. Within each mesh of the layers, we calculated the percentages of the main lithotypes based on the number of pixels corresponding to the lithotype colours, using the automated procedure described in Section 2.3. The results, presented as histograms in Fig. 6b, highlight that Layer 1 contains more than 50% clay-silts in both sections over the entire length of the sections. Furthermore, Layer 2 shows a higher sand content between the line distances D1 and D3 (at 0-3,000 m of section length). In contrast, clay-silts increase between D4 and D6 in the shallow section (3,000-6,000 m of section length) and between D4 and D5 in the deep section (3,000-5,000 m of section length). This lithotype predominates at greater distances in both sections.

Afterwards, we utilize the depth-porosity curves documented in the literature to estimate the sediment compaction and the corresponding reduction in porosity across the section lithotypes. As shown in Fig. 6c, Sclater and Christie (1980) identify an exponential function to model the compaction behaviour of sands (Curve-1 in Fig. 6c), while Baldwin and Butler (1985) define a power-law function to describe the behaviour of shales less than 200 m thick (Curve-2 in Fig. 6c). Furthermore, we adopt an exponential function (red curve in Fig. 6d) introduced by Maselli et al. (2010) to fit the indirect porosity data of shallow muddy layers from a borehole located in the central Adriatic Sea (cloud of grey dots in Fig. 6d). The initial porosity ($\Phi_0$) that appears in the Maselli et al. (2010) function (see Fig. 6d) has been adjusted to better fit the limited porosity data (black dots in Fig. 6d) provided by Cortellazzo et al. (2001). This adaptation resulted in the user-defined curve in Fig. 6d (green line), hereafter referred to as Curve-3. The latter has been obtained using an optimization procedure aimed at minimizing the distance between the two porosity datasets available for the Po Delta area (represented by the black dots in Fig. 6d). This approach allows us to establish a confidence interval for the Maselli curve (represented by the yellow shadow in Fig. 6d) that can be applied to the Po Delta area for the shaly lithotype. This interval ranges from the shaly Curve-2 of Baldwin and Butler (1985) to our optimized (fitted) Curve-3.

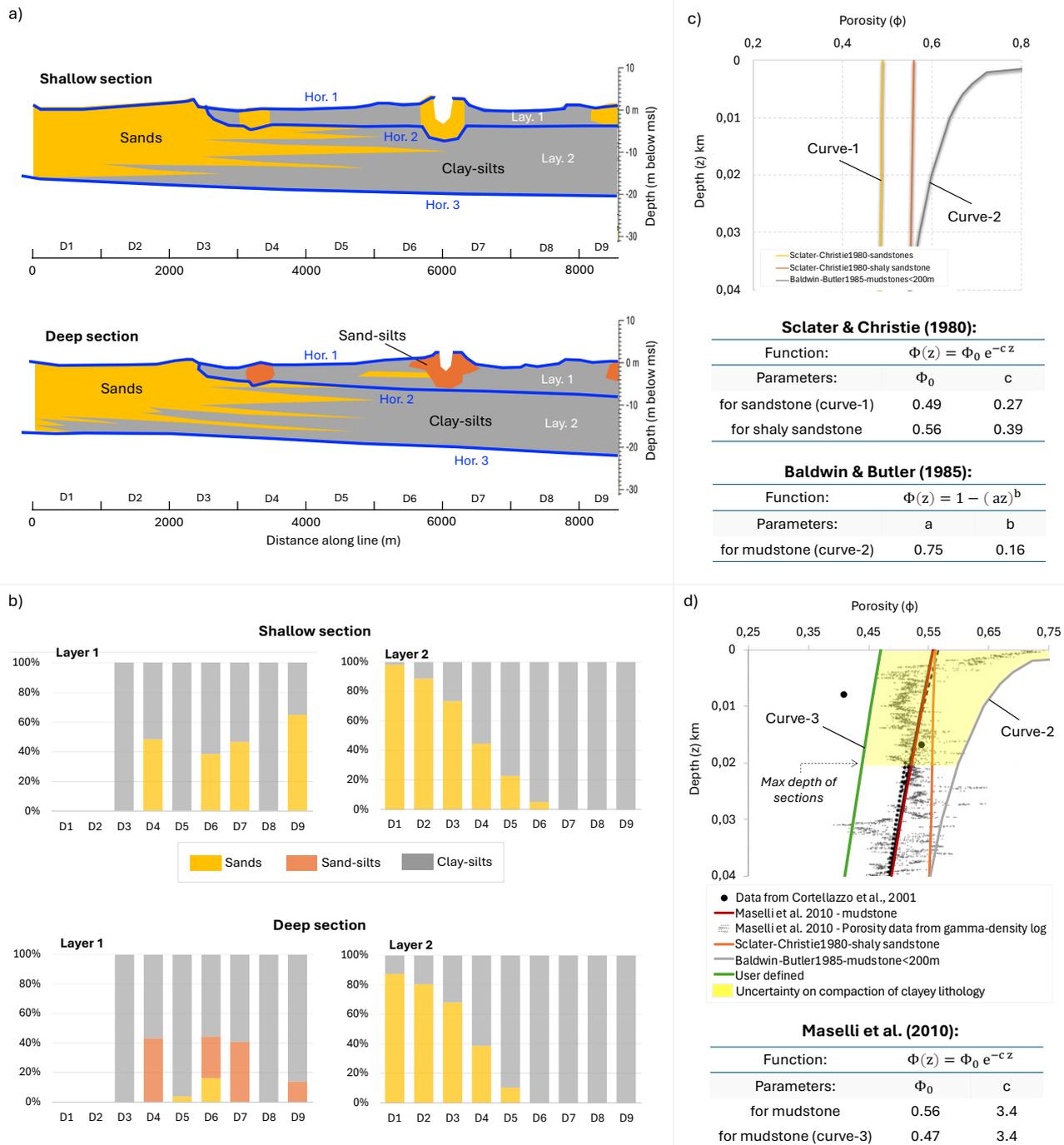

*Figure 6. Lithological content of shallow and deep sections together with porosity-depth curves describing the compaction behavior of different lithotypes. Specifically, the figure includes a) simplified representations of the lithology found in the section images; b) histograms showing the percentages of each lithotype; c) porosity-depth functions derived from the literature and based on global datasets; and d) porosity-depth functions*

*matching local data for clayey lithologies. Labels D1-D9 indicate the line distances on a regular mesh of 1,000 m, used to discretize the section and quantify the percentage of each lithotype (see text for further details).*

Finally, based on the lithotype percentages identified in each layer along the line distance (Fig. 6b), Layer 1 can be characterized by a clayey lithotype compaction behaviour over the entire length of the section, represented by a range of curves between Curves 2 and 3. Layer 2 can be characterized by Curve-1 in the western part of the line (D1-D4) and Curves 2 and 3 in the central-eastern part of the line (D3-D9), allowing for uncertainty in the compaction behaviour as well. In general, both shallow and deep interpretations yield similar qualitative results for Layer 1, except for the most distant D9, and Layer 2. The latter shows a decreasing trend in the percentage of sand lithotype with increasing distance. The most significant discrepancies occur in D4, where the percentage of sands nearly matches that of clay-silts in the shallow section and sand-silts in the deep section. The presence of sands would imply a small variation in the porosity curve, whereas an increase in clay content would lead to a much more significant variation. It is clear that uncertainty can greatly impact the outcomes of any porosity analysis. Quantifying the error still remains a challenging task here, as it depends on epistemic uncertainty (interpreter and measure instrument), but also on aleatory uncertainty (progradation phenomenon).

Concerning the paleo-sea level (PSL), maximum and minimum curves were assigned to layer ages based on global paleo-sea level data (Lambeck et al., 2014) and studies specific to the North Adriatic Sea area (Trincardi et al., 2010; Lambeck et al., 2011). The first curve does not account for glacio-hydro isostatic effects (Fig. 7a) and may be appropriate for the Po Delta area, as in this region the impact of the isostatic response to deglaciation after the Last Glacial Maximum is reduced (e.g., Carminati & Di Donato, 1999). In contrast, the second curve includes the glacio-hydro isostatic correction (Fig. 7a). Due to the discrepancies between the two curves, the error can be evaluated using a linear interpolation technique with the layer ages as query points. The bottom plot in Figure 7a shows that it remains below 2 m and approaches zero (no error) over time as both curves converge to the reference sea level of 0.

Regarding the paleo-water depth (PWD), there are significant differences in the description of lithologies and paleo-environments at both the local scale (i.e., borehole) and the broader regional scale presented in the surface and subsurface interpretations of Sheet 187 and its accompanying notes (Cibin & Stefani, 2009). For example, a single layer may be described as a sediment succession deposited in prodelta and transition to a marine environment according to the subsurface interpretation, while at the same time, it may be defined as sedimentary strata indicative of deltaic and coastal deposits, as suggested by a borehole in the same area described in the accompanying notes.

Concerning the PWD definition, Horizon 3, which corresponds to the base of the prograding sequence, is reconstructed by considering the evolution of bathymetric profiles in the pro-delta zone over the last few centuries (Correggiari et al., 2005), while integrating these data with paleo-environmental insights from lithological descriptions and prograding thicknesses. We construct two scenarios based on the ancient profiles presented in Correggiari et al. (2005), assuming that the basal prograding shape remains unchanged over time (Fig. 7b). The first scenario indicates a water depth ranging from 23.5 m to 25 m (deeper scenario), while the second reflects a shallower depth between 17.5 m and 25 m. The reconstruction of Horizon 2, which corresponds to the top of the prograding stack, follows a similar procedure, but additionally accounts for the reduced thickness of the overlying layer (Layer 1). Two main trends are established, indicating water depths ranging from -0.5 m to 6 m. Importantly, a water depth of 0 m corresponds to the coastline identified in the outcropping deposits and surveyed in the shallow interpretation of Sheet 187. Additionally, the errors shown in red in Figure 7b exhibit an opposition trend for the base and the top of the prograding stack (horizons 3 and 2, respectively). For the base, the error aligns with the declining trend of the associated paleo-bathymetry; it remains consistently elevated (above 5 m) over a short distance but approaches 0 m as the distance increases. Conversely, the error related to the paleo-bathymetry at the top of the prograding stack is zero up to 3,000 m of line distance, then rises between 3,000 and 4,000 m and is characterized by a nearly constant value at greater distances. In this scenario, the error range is significantly smaller than in the previous case, peaking at a maximum of 1 m.

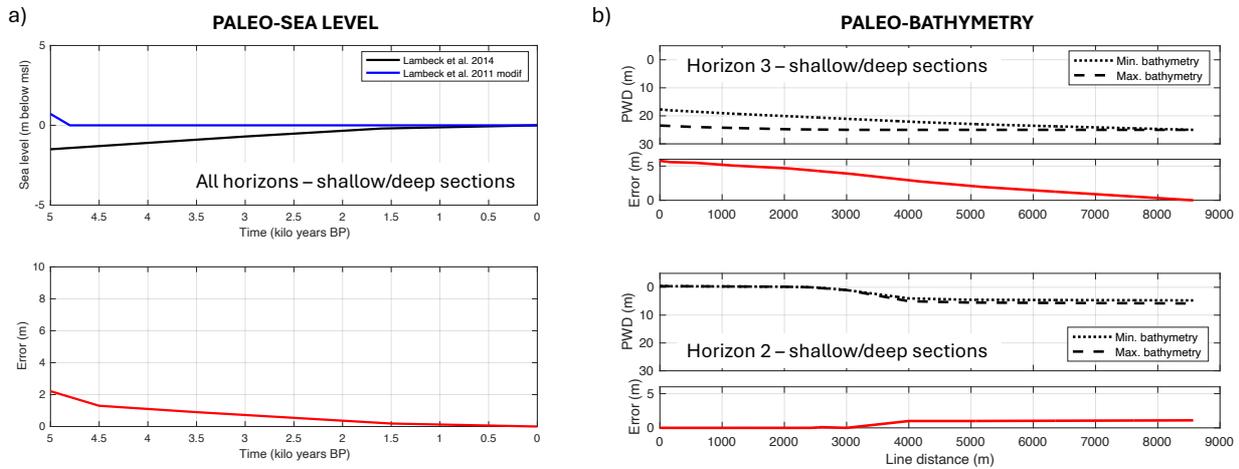

Figure 7. Errors associated with reconstructed: a) paleo-sea levels, b) paleo-water-depths. The red lines in the plots...

## 5. Conclusions

The back-stripping technique in geological sciences is often applied in a simplified manner, primarily qualitatively. The methodology proposed here to consistently categorize the errors presented here addresses geometric aspects of the models, including the availability of different geological interpretations, updated stratigraphic interpretations, and information on eroded thicknesses, as well as stratigraphic and chronological relationships, such as the age of horizons and the geological and environmental conditions, such as eustatic curves. Finally, an alternative method based on geometric analysis of the stratigraphic combinatorics and lines, applied on the original points from relative sea level variations, introduces both qualitative and quantitative treatments as new elements in the application of the back-stripping technique in geological, geodynamic, and environmental fields. Importantly, the research demonstrates how such techniques can be incorporated as error quantification into numerical back-stripping procedures, such as those available in Matlab (e.g., 2021), are widely used in regional studies related to the formation and paleo-bathymetry reconstruction for environmental and climatic integrated error analysis.

The results highlight the applicability of error analysis even in regions with significant data uncertainty. The geological section analyzed is characterized by thin sedimentary layers, underscoring the importance of managing errors and associated uncertainties. The final Figure 8 shows the calculated error values for some of the analyzed categories. The main findings of this analysis are that: 1) the calculated error varies along with the associated probability value; for example, the equiprobability of the error occurring within a specified maximum-minimum range; 2) the magnitude of the calculated error can vary significantly, ranging from metre-scale to millimetre-scale values; 3) the overall effect on the basal depth (BD) and rate of subsidence model is dependent on these errors, as they directly affect the final results. Although present in all cases, differences between shallow and deep interpretations do not appear to have a significant effect on errors arising from the resampling procedure, lithotype attribution, selection of depth-porosity curves, and the thickness of eroded sediment. In the numerical calculation model (see equation 1 in Section 2.1), the BD parameter is directly dependent on the values of the paleo-water depth (PWD) and paleo-sea level (PSL). In contrast, the actual thicknesses affected by the calculated errors are integrated into the decompacted thickness (DT) calculation module and therefore do not directly influence the BD. The largest amount of error is given by the arbitrary extension of Horizon 3, which impacts the thickness of Layer 2. Although this value exceeds the error related to the PWD reconstruction of Horizon 3, it does not directly affect the final result on the BD, because it is involved in the modelling calculation step that combines layer thickness and depth-porosity curves to estimate the decompacted thickness.

In summary, we have attempted to provide a comprehensive explanation of the error classifications introduced, and the standard numerical techniques employed in the quantitative error analysis. In future research, we plan to calculate natural subsidence rates through modelling to assess how errors impact and influence the uncertainty associated with the final results, thus improving our ability to manage and address uncertainty in the outcomes.

**Acknowledgments**
Special thanks to Prof. Simonetta Filippi of the Engineering Department (UCBM, Rome, Italy) for her scientific support during the finalization of the paper.